\documentclass{article}

\usepackage{graphicx}
\usepackage{authblk}
\usepackage[caption=false]{subfig}
\usepackage{amsthm}
\usepackage{amsmath}
\usepackage{amssymb}
\usepackage[margin=1.25in]{geometry}
\usepackage{color}
\usepackage{morefloats}
\usepackage{epstopdf}

\theoremstyle{definition}
\newtheorem{mydef}{Definition}

\usepackage{hyperref}

\begin{document}

\title{Predicting Node Degree Centrality with the Node Prominence Profile}

\author[1]{Yang Yang}
\author[1]{Yuxiao Dong}
\author[1]{Nitesh V. Chawla\thanks{Corresponding Author}}
\affil[1]{Interdisciplinary Center for Network Science and Applications (iCeNSA), Department of Computer Science and Engineering\\
University of Notre Dame\\
Email: yyang1@nd.edu, ydong1@nd.edu, and nchawla@nd.edu}

\maketitle

\begin{abstract}
Centrality of a node measures its relative importance within a network. There are a number of applications of centrality, including inferring the influence or success of an individual in a social network, and the resulting social network dynamics. While we can compute the centrality of any node in a given network snapshot, a number of applications are also interested in knowing the potential importance of an individual in the future. However, current centrality is not necessarily an effective predictor of future centrality. While there are different measures of centrality, we focus on degree centrality in this paper. We develop a method that reconciles preferential attachment and triadic closure to capture a node's prominence profile. We show that the proposed node prominence profile method is an effective predictor of degree centrality. Notably, our analysis reveals that individuals in the early stage of evolution display a distinctive and robust signature in degree centrality trend, adequately predicted by their prominence profile. We evaluate our work across four real-world social networks. Our findings have important implications for the applications that require prediction of a node's future degree centrality, as well as the study of social network dynamics.
\end{abstract}

\section*{Introduction}
Social networks spurred by digital innovations, such as Facebook, LinkedIn, and Twitter, make up an increasingly wide range of diverse human interactions. These social networks are dynamic and evolve over time, wherein new nodes enter a network, new links may form between nodes or old links may diminish between nodes, and a node's centrality may change over time. Thus, the node and the network co-evolve, where the node impacts the network and the network impacts the node, creating an intertwined effect of centrality and relative position of the node~\cite{gross:2008}. As the network evolves, we are interested in knowing the predictability of centrality of a node. Prediction of centrality can lead us to infer influence, importance and/or success of a given individual in a social network. We use the popular degree centrality as a metric in this paper (various studies have found centrality measures to be correlated \cite{rothenberg:1995} \cite{valente:2008}).

Over the last decade, network evolution modeling focused on defining basic mechanisms driving link creation and capturing different properties observed in real networks, such as \text{power-law} degree distribution, small diameter, and clustering coefficient as function of node degree centrality \cite{acm:preferential} \cite{watts:98} \cite{acm:jure2008}. However, the network evolution drives not only the emergence of macroscopic scaling of social networks but also the microscopic behaviors of individuals. The Barabasi-Albert model \cite{acm:preferential} provides a mechanism for the emergence of scale-free property in social networks, where new links are established preferentially to well connected individuals. However, it is also evident that \text{preferential attachment} is not sufficient to reproduce other important features of social networks, and individual's link formation also significantly relies on its neighbors \cite{acm:newman2001, acm:li2010}. The principle of {triadic closure} has been empirically demonstrated to be relevant for several macroscopic scaling laws in the work of \cite{li:2013} \cite{acm:triad1} \cite{acm:triad2} \cite{acm:jure2008} \cite{acm:validation1} \cite{acm:validation2} \cite{acm:validation3}, explicitly or implicitly. The triadic closure mechanism is based on the premise that two individuals with mutual friends have a higher probability to establish a link. These two principles successfully capture the main characteristics of social networks. However \text{preferential attachment} requires global information while {triadic closure} only needs local information \cite{acm:li2010}. Triadic closure captures the notion of relative position of a node. Irrespective of the specific mechanisms that act to drive the emergence of macroscopic scaling of social networks, it is reasonable to ask whether such mechanisms also shape the microscopic behaviors of individuals --- the degree centrality change. Can we effectively predict degree centrality of a node in the future? We find that the current degree centrality is a weak predictor of the future degree centrality. Rather, the degree centrality evolution is an artifact of both the centrality (preferential attachment) of the node and its relative position (triadic closure) in the network, and is a suitable proxy for interpreting the evolution of a network from a microscopic perspective. We define this combination of centrality and position as prominence. A node may become important over time, which may be a result of its individual achievement or neighborhood structure, which represent aspects of preferential attachment and triadic closure respectively. A node is prominent if the links to the node make it visible to the other nodes in the network \cite{acm:prominence, acm:prominence1}. The prominence of a node also depends on the overall structure of its neighborhood. {{This paper develops a methodological framework that characterizes the prominence of an individual by reconciling the trade-offs between {preferential attachment} and {triadic closure}, that is the microscopic level, and develops a model to predict degree centrality of a node in the future.}} We call our framework the {\it Node Prominence Profile (NPP)}. 
\begin{figure}[tp]
	\centering
		\includegraphics[width=2.8in]{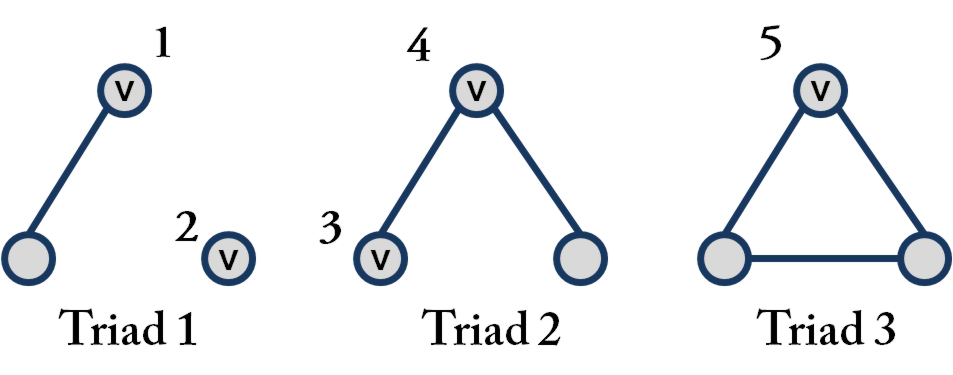}
\caption{Node Prominence Profile. In this figure, we mark $5$ automorphism positions (labeled with $v$) in $3$ triad structures. The node prominence profile of a node $v$, is a vector describing the occurrence frequencies of node $v$ in these $5$ automorphism positions.}
  	\label{fig_tpp_example}
\end{figure}

Formally, the {\it Node Prominence Profile} is defined as follows:
\begin{mydef} {\bf Node Prominence Profile}
{\it Node Prominence Profile} for a node $v$, written as $\text{NPP} (v)$, is a vector describing the occurrence frequencies of node $v$ in five different positions of three isomorphic triad substructures (Figure~\ref{fig_tpp_example}).
\label{def_1}
\end{mydef}
In Figure~\ref{fig_tpp_example} we demonstrate $5$ automorphism positions in $3$ triad sub-structures. These three triad sub-structures were discussed in social balance theory proposed by Heider \cite{heider:1958}. To compute the node prominence profile for an individual node $v$, we just need to find out all triad sub-structures where node $v$ is located; and then we count how many times node $v$ occurs in each automorphism position (see \textcolor{blue}{Supporting Information, 2.3}). In this way a high degree centrality node tends to be located in many triad $1$ and triad $2$ sub-structures, and has high occurrence frequencies in position $1$ and position $4$ correspondingly. Based on the principle of preferential attachment, they are more likely to attach new links in future. Nodes in position $2$ are not necessarily isolated, they just do not have direct links with other two nodes in such a sub-structure (Triad $1$). Clearly nodes in position $2$ have potential to develop new links in future. Driven by triadic closure effect, triad $2$ tends to evolve to triad $3$, thus nodes having high occurrence frequencies in position $3$ are more likely to attach new links. Triad $3$ is a stable sub-structure \cite{heider:1958}, as the propensity of attaching new links for the nodes having high occurrence frequencies in position $5$ is relatively small. Positions in these three triad structures embody both principles --- preferential attachment and triadic closure.

The empirical experiments reveal that {\bf{NPP}} is able to provide more precise prediction of node's future degree centrality over baseline solutions. {\bf{NPP}} is validated on four different social networks. We also demonstrate that the model developed on one social network and predict on another social network (transfer learning), thus demonstrating the generalization capacity of {\bf{NPP}} and confirming that it is effective in capturing the general factors underlying social network evolution impacting the degree centrality of a node.

\section*{Results}
\subsection*{Node Prominence Profile}
As we posited, prominence is not only represented in the node's centrality but also in the node's position in local structure. Much effort has been devoted to measuring the node's centrality, such as {degree centrality}, {Pagerank} \cite{acm:pagerank}, {Betweenness} \cite{acm:betweenness}, and {Closeness} \cite{acm:closeness}. Here we will demonstrate that modeling prominence can lead to a much improved prediction about a node's future degree centrality in the network than modeling current state-of-the-art centrality measures. % degree centrality itself.

\begin{figure}[tp]
	\centering
		\includegraphics[width=5in]{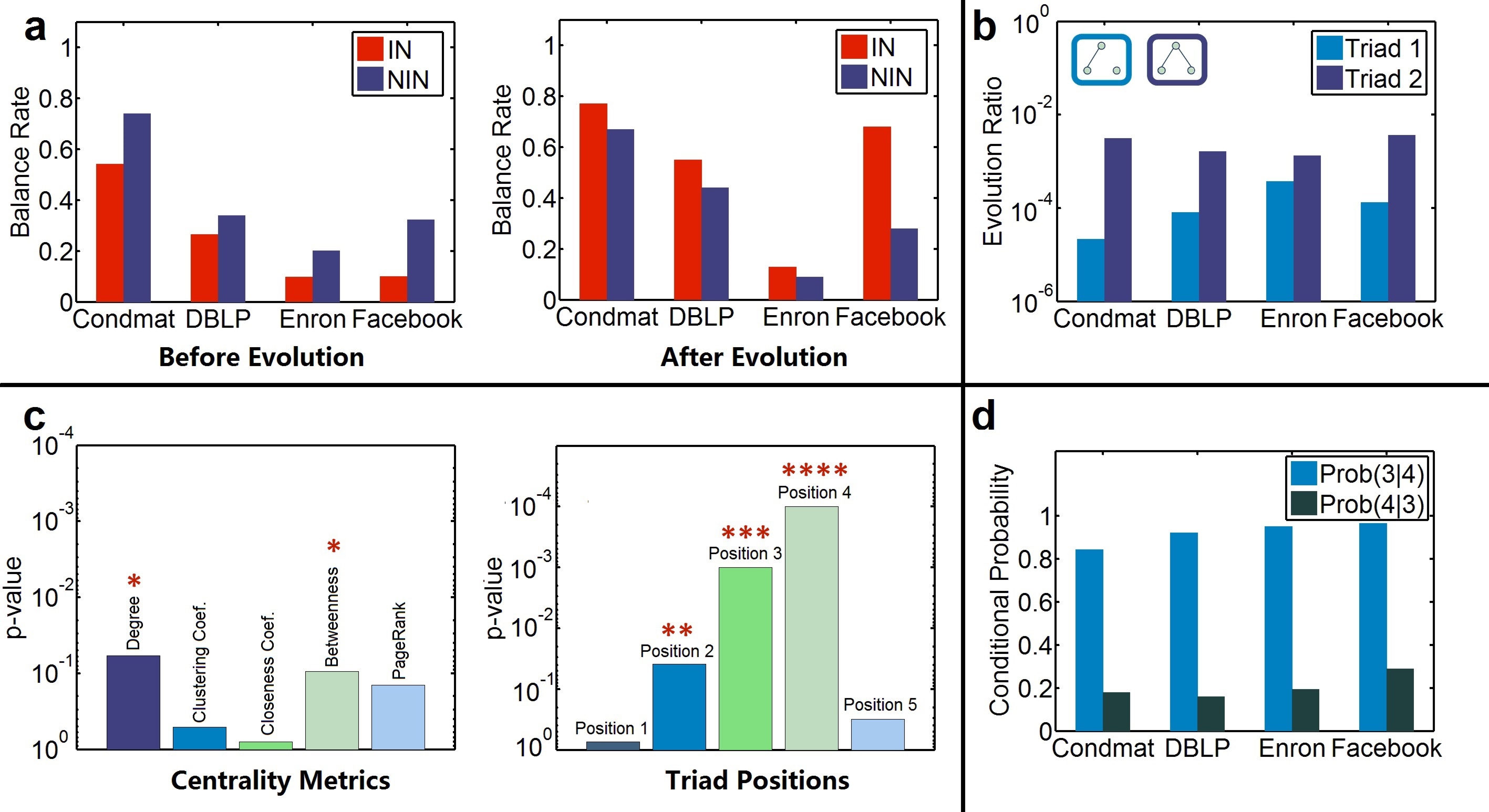}
\caption{Microscopic Prominence Analysis. {\bf (a)} Structural Balance Rate. For the nodes joining the network $G_{t}$ at the same time $t$, based on their degree centrality in the network $G_{t+\Delta t}$ after $\Delta t$ timestamps, we divide them into two sets {\bf Important Nodes} and {\bf Non Important Nodes} (see \textcolor{blue}{Supporting Information, 1.2} for detail). (left) In the network $G_{t}$ we extract the sub-networks of {\bf IN} and {\bf NIN} and calculate their balance rates correspondingly. We observe that the {\bf IN} sub-network has a lower balance rate than the {\bf NIN} sub-network. (right) Similarly in the network $G_{t+\Delta t}$ we extract the sub-networks of {\bf IN} and {\bf NIN}, the {\bf IN} sub-network has a larger balance rate than the {\bf NIN} sub-network. {\bf (b)} Triad Evolution Rate. In four datasets we compute the link formation probability within different kinds of triads, we call it triad evolution rate. We observe that the ``forbidden'' triad ({\it triad 2}) (Figure~\ref{fig_tpp_example}) has much higher probability to form a new link than the disconnected sub-structure {\it triad 1}.(\textcolor{blue}{Supporting Information, 2.3}). {\bf (c)} Significance of Inferring Future Degree Centrality. We consider these centrality measures and positions as predictors of future degree centrality, we show the $p$-value associated with each feature and its corresponding significance level under Wald test (\textcolor{blue}{Supporting Information, 2.3}). {\bf (d)} Position Conditional Probability. We calculated the conditional probability of position $3$ and position $4$ (see Figure~\ref{fig_tpp_example}), $Prob(3|4)$ states the probability that a node shows up in position $3$ given the condition that it is located in position $4$; $Prob(4|3)$ is the probability that a node is located in position $4$ given the condition that it is also in position $3$. We can see that nodes in position $4$ have high probability to be located in position $3$, while nodes in position $3$ have less than $0.3$ probability to occur in position $4$.}
  	\label{fig_micro_prominence}
{
\begin{flushleft}
\scriptsize *: p $<$ 0.1; **: p $<$ 0.05; ***: p $<$ 0.01, ****: p $<$ 0.001.
\end{flushleft}
}
\end{figure}

\paragraph*{Triadic Closure Effect on Degree Centrality Evolution}
The effect of {preferential attachment} on the degree centrality evolution is evident \cite{acm:preferential}. However, the prominence of a node not only includes its centrality but also its position in local neighborhood, and the principle of {\it preferential attachment} is inherently unable to describe node's position in local structure \cite{acm:li2010}. The {triadic closure} principle provides us an alternate solution. We first study the effect of {triadic closure} on the degree centrality evolution. The quantity of {triadic closure} (or structural balance) can be defined as below \cite{acm:newman2001} (\textcolor{blue}{Supporting Information, 2.2}):
\begin{equation}
\text{balance rate} = \frac{3 \times \text{number of closed triads}}{ \text{number of connected triads}} 
\end{equation}
By studying the sub-networks of important (high degree centrality) or non-important (low degree centrality) nodes (based on {Pareto Principle} \cite{acm:pareto}, we partition nodes into important and non-important nodes based on their degree centrality, denoted as {\bf IN} and {\bf NIN}, see \textcolor{blue}{Methods} and \textcolor{blue}{Supporting Information, 1.2}), we observe that initially the sub-network of future important (having high degree centrality in future) nodes has a lower balance rate than the sub-network of future non-important nodes, but the former sub-network evolves to form a more balanced topology (Figure~\ref{fig_micro_prominence} (a)). There are several implications: 1) there exist connections between the {triadic closure} and the degree centrality evolution. In addition, new links are more likely to form between nodes located in an unbalanced sub-network; 2) The initial sub-network where future important nodes are located is more imbalanced than that of future non-important nodes, thus position of node can be indicative of its future degree centrality. These findings are demonstrated to be statistically significant at 95\% confidence even if we scale the threshold value of important nodes, such as $10\%$, $30\%$ and $50\%$ (see \textcolor{blue}{Supporting Information, 2.2}). This implies the possible effect of {triadic closure} on both the node's degree centrality and its position in local neighborhood.
%insert the figure

As suggested in the principle of {triadic closure}, a ``forbidden'' triad \cite{acm:newman2001} ({\it triad 2} in Figure~\ref{fig_tpp_example}) is more likely to attach new links. In order to demonstrate that position is crucial for the degree centrality evolution, we provide the evolution ratio of two types of triads in Figure~\ref{fig_tpp_example}. {For a triad structure, if there are new links attached, then we say this triad structure evolves. And for a specific type of triads (i.e., triad $1$), we calculate how many percentage of them evolve and denote that as the evolution ratio (see \textcolor{blue}{Supporting Information, 2.2}). We can see that the ``forbidden'' triad ({\it triad 2}) has much higher probability to attach a new link than a disconnected sub-structure {\it triad 1} (Figure~\ref{fig_micro_prominence} (b)). This implies that nodes in different genres of triads have different probabilities to develop degree centrality. This leads us to an important conclusion: the positions of nodes in sub-structures determine their future orbits in both essential prominence elements for describing degree centrality evolution. Our methodological framework called, the Node Prominence Profile (Definition~\ref{def_1}), incorporates these insights in the modeling for the node's prominence (Figure~\ref{fig_tpp_example}; (\textcolor{blue}{Supporting Information, 2.3})).

\paragraph*{Positions in Triad Structure}
In Figure~\ref{fig_tpp_example} we enumerate all possible five positions in the triad sub-structures described in Definition~\ref{def_1}. We observe that the position of a node within its local structure is related with its degree centrality evolution. For instance in Figure~\ref{fig_micro_prominence} (b) nodes in ``forbidden'' triad are more likely to attract new links (\textcolor{blue}{Supporting Information, 2.3}). Thus, different positions of a node in corresponding triads should have distinct descriptive power of degree centrality evolution.

In Figure~\ref{fig_micro_prominence} (c) we provide the significance of different centrality measures and position incidence values in indicating node's future degree centrality ({\bf IN} or {\bf NIN} in future).

We observe (see Figure~\ref{fig_micro_prominence} (c)) that the centrality measures are not performing well in describing a node's future degree centrality except degree centrality and betweenness centrality. At the same time several position incidence values are significantly better in inferring a node's latent degree centrality. In the experiment the sets of ${\bf IN}_{t+\Delta T}$ (nodes with high degree centrality at time $t+\Delta T$) and ${\bf NIN}_{t+\Delta T}$ (nodes with low degree centrality at time $t+\Delta T$) (\textcolor{blue}{Supporting Information, 1.2}) are labeled based on the {degree centrality} (Figure~\ref{fig_micro_prominence}). We note that the {\it degree centrality} metric itself does not have the most significant correlation with node's future degree centrality when $\Delta T$ is large (\textcolor{blue}{Supporting Information, 1.2}). Based on these observations, we have several conclusions: 1) different positions have different power in describing node's future degree centrality; 2) three of them are much better than centrality measures themselves. To {\bf summarize}, even though the centrality measures are demonstrated to be good at centrality (relative importance in the network) quantification, they are not powerful enough to depict the node's future degree centrality. This is because, the preferential attachment is not the only origin underlying the social network dynamics \cite{acm:newman2001} \cite{acm:jure2008} \cite{acm:li2010}. Additionally we can observe that positions in triad structures embody both principles---preferential attachment and triadic closure. Triad position $1$ and $4$ reflect the effect of preferential attachment, while triad position $3$ manifests the triadic closure principle. This confirms our propositions made above and provides a possible way to balance the effects between preferential attachment and triadic closure and model two essential elements of node's degree centrality effectively.

As {\it triadic closure} principle suggests, for the unclosed triad ({\it triad $2$}) new links are attached between nodes in position $3$, however we have observed that nodes having high occurrences in position $4$ are more likely to have high degree centrality in future. One possible reason underlying such phenomenon is, nodes in position $4$ have higher attraction to links ({\it preferential attachment}). However in Figure~\ref{fig_micro_prominence} (c) we already identify that {\it degree centrality} does not have a comparable performance as the position $4$. To further investigate this, we calculated the conditional probability of position $3$ and position $4$, $Prob(3|4)$ states the probability that a node shows up in position $3$ in one triad given the condition that it is located in position $4$ in a different triad; $Prob(4|3)$ is the probability that a node is located in position $4$ given the condition that it is also in position $3$. In Figure~\ref{fig_micro_prominence} (d) we can see that nodes in position $4$ have extremely high probability to be located in position $3$ (close to 1.0), while nodes in position $3$ have less than $0.3$ probability to be in position $4$. This means, nodes in position $4$ are influenced by both mechanisms of {\it preferential attachment} and {\it triadic closure}, while nodes in position $3$ are mainly affected by the {\it triadic closure} principle. This explains why position $4$ has higher significance level than position $3$. This also reflects an {\bf important} property of these positions in triad structures---they combine the two well known social principles (i.e. {\it preferential attachment} and {\it triadic closure}).

\subsection*{Prominence: Centrality and Position}
In order to demonstrate that prominence is not only represented in the node's centrality but also in the node's position in local structure, we provide a detailed investigation into their interaction from the perspective of {\it influence events} and provide the evidence that the {\bf NPP} is able to model both centrality and position information. In order to validate their connections, we define {\it link influence} (Figure~\ref{fig_triad_event}) between two nodes $u$ and $v$ (\textcolor{blue}{Supporting Information, 2.4}).
\begin{mydef}
For a given node $u$ in the time-varying network $G = (V, E, T_{V},T_{E})$ (see \textcolor{blue}{Supporting Information, 1.2 Definition 3}), $u$ is said to have a {\bf link action} on node $w$ at time $t$ if $(u, w) \in E$ and $t \in T_{E} (u, w)$. $T_{V}$ is the log of nodes joining timestamps, while $T_{E}$ is the log of edge formation timestamps.
\end{mydef}
Additionally we provide the definition of the {\it link influence} of node $u$ on its neighbor $v$ as follows:
\begin{mydef}
A node $u$ is said to have a {\bf link influence} on its neighbor $v$ {\it iff}: 1) there is a link action of node $u$ with another node $w$ at time $t$; 2) there exists a link action of node $v$ with node $w$ at time $t'$; 3) $min(T_{E}(u,v)) < t < t'$ and $t' - t < \sigma$
\end{mydef}
The $\sigma$ is the average action delay between two nodes $u$ and $v$. An example of {\it link influence} is presented in Figure~\ref{fig_triad_event}.
%insert the figure
\begin{figure}[htp]
	\centering
		\includegraphics[width=2.5in]{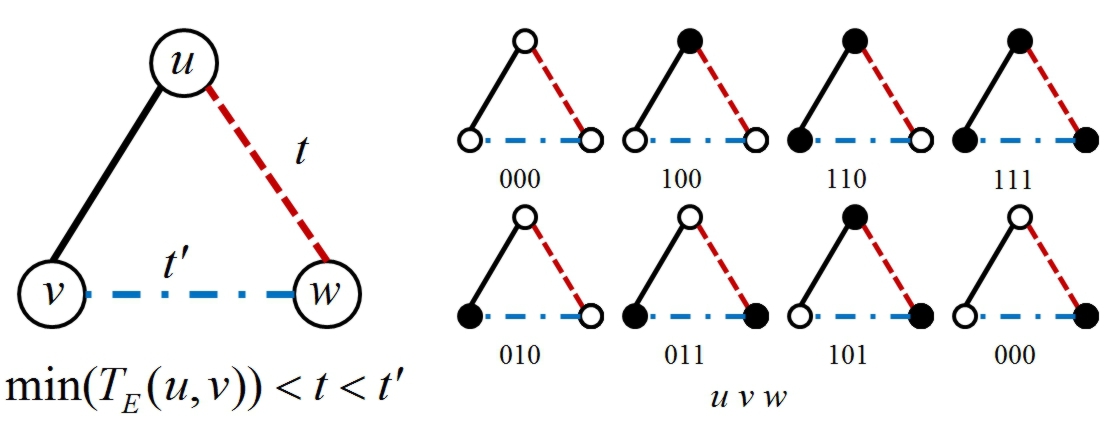}
\caption{Link Influence Events. On the left side, we demonstrate the link influence of node $u$ on its neighbor $v$. Node $u$ has a link action with node $w$ at time $t$, and node $v$ has a link action with node $w$ at time $t'$. And the gap between $t$ and $t'$ is smaller than a threshold $\sigma$. $\sigma$ is the average action delay between two nodes $u$ and $v$. On the right side, we enumerate all possible kinds of link influence events if nodes in the network are partitioned into important nodes and non-important nodes. The three digits encode the degree centrality status of the three nodes, $u$, $v$, and $w$, `1' indicates an important node (high degree centrality) and `0' indicates a non important node (low degree centrality). Thus there are $8$ kinds of link influence events.}
  	\label{fig_triad_event}
\end{figure}

We divide the nodes into two groups (important nodes and non important nodes), as considered in Figure~\ref{fig_micro_prominence}. As shown, in Figure~\ref{fig_triad_event}, we partition the {\it link influence} event into $2{^3} = 8$ categories based on nodes' degree centrality; `1' indicates an important node (high degree centrality) and `0' indicates a non important node (low degree centrality). In Figure~\ref{fig_triad_event_stat} we provide the distribution of several patterns, and we observe that: 1) $|1XX| > |0XX|$ and $|X1X| > |X0X|$ ($|1XX|$ is the number of link influence events where node $u$ is an important node, Figure~\ref{fig_triad_event}), this means important nodes have much higher probability to have {\it link influence} on their neighbors, and it also validates the principle of {\it preferential attachment}; 2) additionally $|XX0| > |XX1|$, non-important nodes play an important role to transfer {\it link influence}; 3) $|11X| > |00X|$, this states that {\it link influence} is more likely to happen between important nodes; 4) $|10X| \approx |01X|$, if {\it link influence} occurs among important nodes and non-important nodes, then important nodes and non-important nodes have the same chance to initiate the influence. These patterns persist in four different real-world networks and are proved to be statistically significant (see \textcolor{blue}{Supporting Information, 2.4}). Thus, this further implies that interactions between degree centrality and position (link formation leads to the change of node's position) are common in social networks.

\begin{figure}[!h]
	\centering
		\includegraphics[width=3.2in]{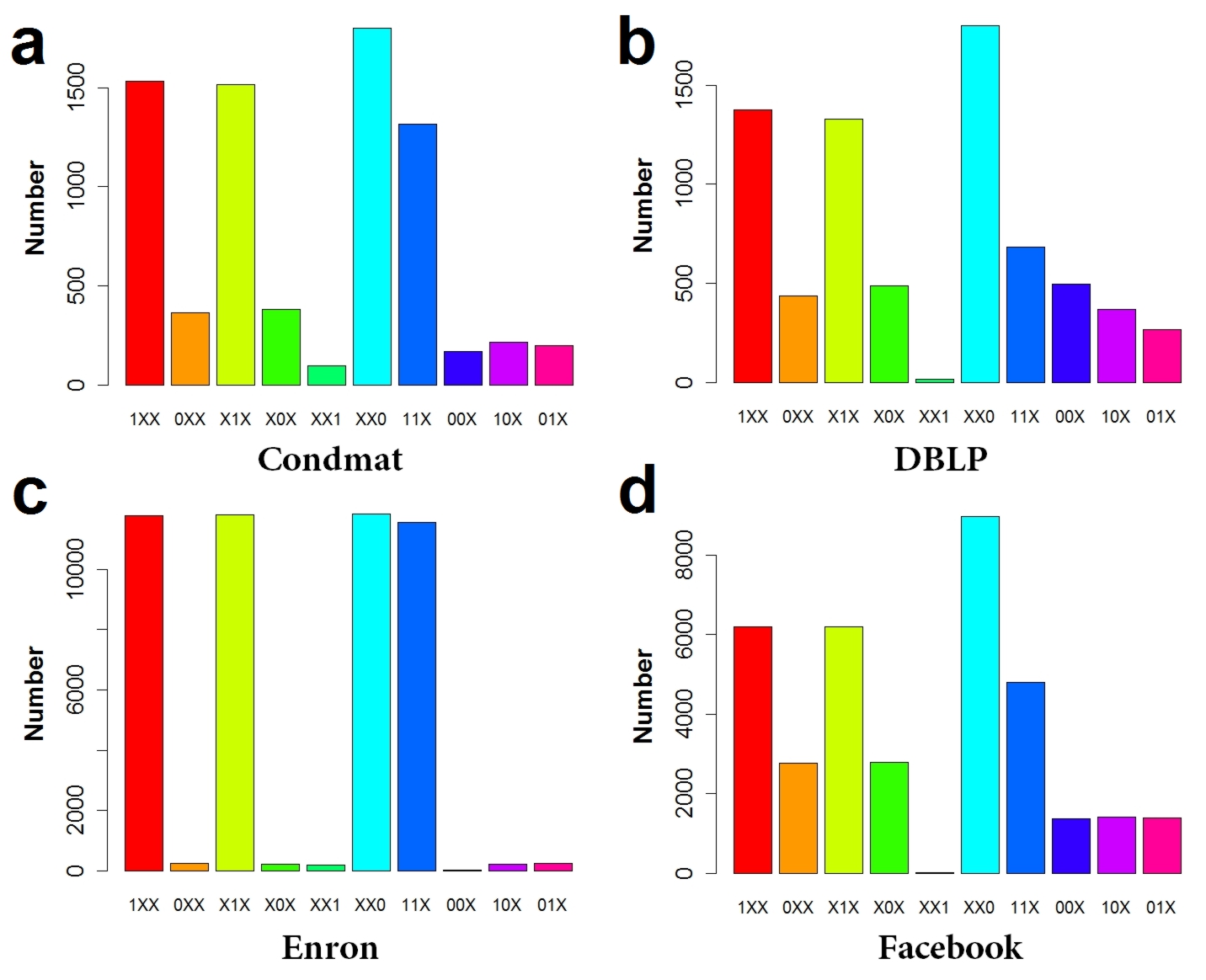}
	\caption{Degree Centrality Status vs. Link Influence Event. `1' indicates an important node (high degree centrality), `0' indicates a non important node (low degree centrality), and `X' indicates that the code is either `1' or `0'. We observe similar patterns in four different real-world social networks.}
\label{fig_triad_event_stat}
\end{figure}

\subsection*{Prediction of Future Degree Centrality}
The {\bf NPP} method described above allows us to investigate two aspects of node's prominence. We present empirical analysis to validate that our method is able to more effectively predict a node's future degree centrality than the state-of-art methods.
\paragraph{Inferring Future Degree Centrality}
In Table~\ref{performance_metric1}, we provide an empirical comparison of the performance for the problem of degree centrality prediction in terms of AUROC (Area under the ROC curve) and AUPR (Area Under the Precision-Recall Curve) (see \textcolor{blue}{Methods}). Our approach, {\bf NPP}, outperforms three baseline methods in terms of {\it AUPR}, and has better or comparable performance in terms of {\it AUROC}. {\bf All} method includes existing state-of-the-art centrality measures, which is described in \textcolor{blue}{Supporting Information, 3.1}. We have several {\bf conclusions}: 1) the principle {\it preferential attachment} is just one dimension of mechanisms underlying the nodal degree centrality evolution; 2) the trade-offs between {\it triadic closure} and {\it preferential attachment} are well balanced in {\bf NPP} and then it achieves better performance in the degree centrality prediction task. Additionally our model NPP is also able to predict nodes' future degree centrality and yield better performance than the state-of-art methods (see \textcolor{blue}{Supporting Information, 3.3}). This further confirms the correctness and effectiveness of our methodology.

\subsection*{Generality of Node Prominence Profile: Prediction Across Datasets}
\label{sec_generalization}
We have demonstrated that the {\bf NPP} has a stronger generalization capacity than state-of-the-art centrality measures in predicting future degree centrality. To be rigorous, we now ask: are these features powerful enough to transfer learning from one social network to another? That is, can a model developed on one social network effectively predict for another social network? If our framework is able to generalize across datasets, then it will further demonstrate that our framework captures the essential principles of degree centrality evolution.

\begin{figure}[htp]
	\centering
		\includegraphics[width=3.9in]{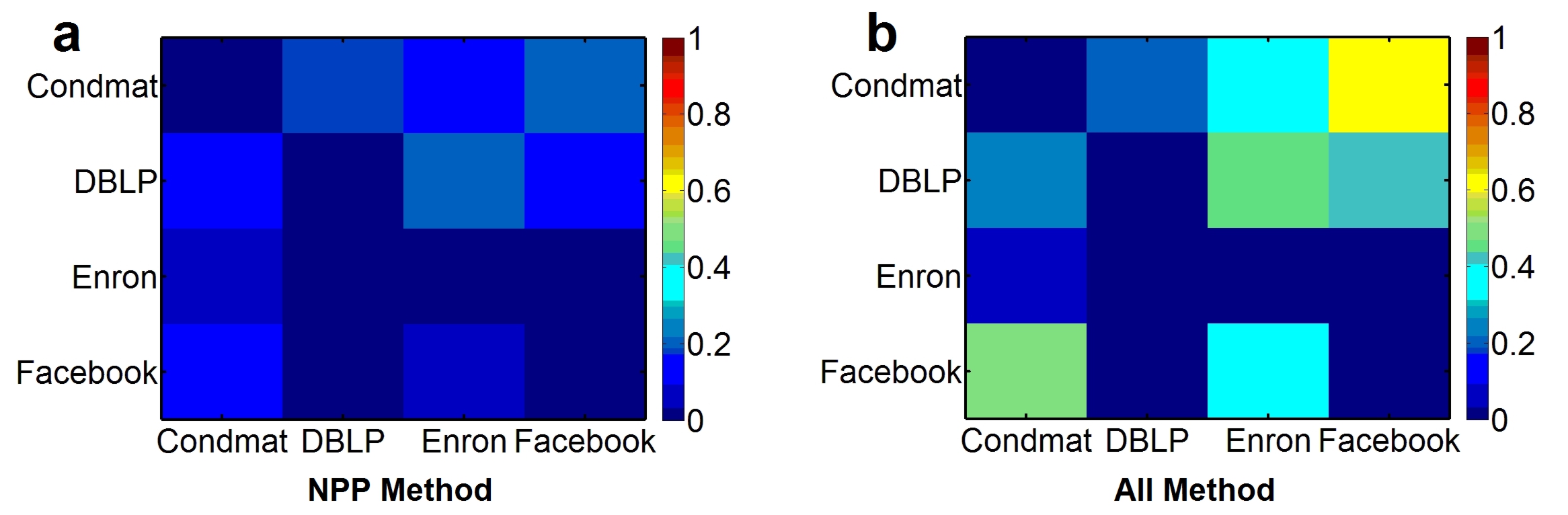}
	\caption{Generalization Performance Loss in AUPR (Degree Centrality Prediction). Different from the learning task performed in single dataset, the training set is extracted from one dataset and the prediction (testing set) is made on another dataset. AUPR, area under precision-recall curve. The AUPR score is more sensitive than AUROC in reflecting the difference of prediction \cite{acm:linkprediction6}. In order to demonstrate stability of generalization, we use AUPR for the performance evaluation. The detail of {\bf NPP} and {\bf All} methods can be found in \textcolor{blue}{Supporting Information, 3}. {\bf All} method includes existing centrality measures, which is described in \textcolor{blue}{ Supporting Information, 3.1}. Each element represents the performance reduction compared with the regular learning results (i.e., training and testing on the same dataset). We can observe that the performance reductions of {\bf NPP} method are mostly less than 20\%, while the performance reduction of {\bf All} method can achieve about 60\%.}
\label{table_general_across}
\end{figure}

\paragraph*{Generalization of the Degree Centrality Prediction}
In Figure~\ref{table_general_across}, we provide the transfer learning (transfer of learning is usually described as the process and the effective extent to which past experiences (trained model) affect performance (prediction) in a new situation or data different from the one that the model was trained on) results for {\bf All} model and {\bf NPP} model. Each pair of generalization is trained on the row dataset and evaluated on the column dataset. Transfer learning generally leads to the loss in performance. In Figure~\ref{table_general_across} we provide the performance loss of transferred learning compared with the performance of non-transfer learning (where training and testing are conducted on the same dataset). Thus, the diagonal entries all have performance loss as zero.

There are several observations. We observe that the {\bf NPP}'s performance degrades remarkably less than the {\bf All} method in most cases. This indicates that the prominence profile of node captures principles that are more generic than the state-of-the-art centrality measures, and this still holds even if the generalization is across different domains of networks. This further confirms that the prominence profile is a general cross-domain property for the degree centrality evolution analysis. In conclusion, the prominence profile is notably more generic across different domains of networks, and the state-of-the-art centrality based method is more particular to a specific dataset.

% references section
\section*{Discussion}
We analyzed several principles/mechanisms underlying the network evolution, mainly focusing on two essential elements of the node's prominence: centrality and position. We demonstrated that the position of a node in a local structure is strongly indicative of the degree centrality progression in the social network. Building on this observation, we developed a prediction method referred to as {{\bf NPP}}. We empirically demonstrated the effectiveness of {{\bf NPP}} by demonstrating improvement in performance over the state-of-art methods for the problem of degree centrality prediction in four different datasets. We further established the generalization capacity of our methods under a transfer learning scenario --- we learned the classifier on one social network (using the proposed features) and tested on another social network. The performance trends clearly showed that our approach is able to capture essential properties or features underlying degree centrality evolution, which are general across different domains of social networks.

Our methodology ({\bf{NPP}}) is validated to optimize trade-offs between essential dimensions of network evolution ({\it preferential attachment} and {\it triadic closure}). Therefore, it is not surprising that, as a consequence, our approach yields accurate and generic performance in predicting node's future degree centrality. Our method can be effectively used in a variety of applications that rely on inferring a node's importance in the future, as captured by centrality measures. In summary, we have developed a new perspective for growth in degree centrality of a node in a social network and developed a general purpose feature vector that can be used by different machine learning algorithms across different social networks.

\section*{Methods}
\label{sec_methods}
\noindent{\bf Data Description}\\
In this paper we examine our approaches and perform our analysis on four social networks. The {\it Condmat} network \cite{acm:linkprediction2} is extracted from a stream of 19,464 multi-agent events representing condensed matter physics collaborations from 1995 to 2000. Based on the {\it DBLP} dataset from \cite{acm:dblp} we attach timestamps for each collaboration and choose 3,215 authors who published at least 5 papers. {\it Enron} dataset \cite{acm:enron} contains information of email communication among 16,922 employees in Enron Corporate from 2001.1.1 to 2002.3.31. The {\it Facebook} dataset is used by Viswanath et al. \cite{acm:facebook}, which contains wall-to-wall post relationship among 11,470 users between 2004-10 and 2009-01.

\noindent{\bf Important Nodes and Non Important Nodes}\\
On a global level, important nodes have intrinsically higher strength of impact than others due to the network topology. Through our study, we have found that a small number of nodes occupy large portion of network resources. For example, in \textcolor{blue}{Supporting Information, 1} we observe that top 20\% (ranked by {PageRank}) nodes occupy about 80\% {\it PageRank} centrality in DBLP network. This satisfies {\it Pareto Principle} (also known as 80-20 rule) \cite{acm:pareto}. To better understand and model the effects of network evolution on node's prominence, we partition nodes into two sets {\it important nodes} and {\it non-important nodes}.
{\bf Important Node}: In a network $G=(V, E)$ a node $v$ is a {\it important node} under centrality measurement $\mathbb{M}$ if and only if $ \frac{| \{ u | \mathbb{M}(u) \leq \mathbb{M}(v) \}|}{|V|} \geq 0.8$. {\bf Non-Important Node}: In a network $G=(V, E)$, a node $v$ is a {\it non-important node} under centrality measurement $\mathbb{M}$ if and only if $ \frac{| \{ u | \mathbb{M}(u) > \mathbb{M}(v) \}|}{|V|} \geq 0.2$. In following sections we denote the set of {\it important nodes} as ${\bf IN}$ and the set of {\it non-important nodes} as ${\bf NIN}$.

\noindent{\bf Evaluation Methods}\\
We employ AUROC and AUPR to evaluate the performance of the predictions tasks in this work. The information of associated evaluations metrics are as below:\\
\textit{ROC:} The receiver operating characteristic (ROC) represents the performance trade-off between true positives and false positives at different decision boundary thresholds\cite{mason:2002,fawcett:2004}.\\
\textit{AUROC:} Area under the ROC curve.\\
\textit{Precision-recall Curve:} Precision-recall curves are also threshold curves. Each point corresponds to a different score threshold with a different precision and recall value \cite{davis:2006}. \\
\textit{AUPR:} Area under the precision-recall curve.

\noindent{\bf Prediction Experiment Settings}\\
For the prediction of future degree centrality, we use Bagging with {\it Logistic Regression} as the supervised learning model. Bagging \cite{bagging:2011} is a machine learning ensemble meta-algorithm designed to improve the stability and accuracy of machine learning algorithms used in statistical classification and regression. The bagging method reduces variance and helps to avoid over-fitting, which is usually applied to many types of machine learning methods. Our goal here is to evaluate the utility of additional information imputed by us in the feature vector versus the quality of a learning algorithm. In our experiment we only allow methods to observe features of nodes in a short duration after nodes joining the network, for example, for Condmat and DBLP we only use the first year data of new arriving nodes and for Enron and Facebook we only use the first month data of new arriving nodes. We classify the nodes in to {\bf IN} and {\bf NIN} using {\it degree centrality} (\textcolor{blue}{Supporting Information, 3}).

In order to demonstrate the generality of our framework, we perform the transfer learning for the Degree Centrality Prediction problem. Each pair of generalization is trained on the one dataset and evaluated on another dataset by Bagging with {\it logistic regression} (for example, trained on Condmat and evaluated on DBLP) (\textcolor{blue}{Supporting Information, 4}).

%ACKNOWLEDGMENTS are optional
\section*{Acknowledgments}
Research was sponsored by the Army Research Laboratory under Cooperative Agreement Number W911NF-09-2-0053, and by the grant FA9550-12-1-0405 from the U.S. Air Force Office of Scientific Research (AFOSR) and the Defense Advanced Research Projects Agency (DARPA)

\section*{Author Contributions Statement}
YY and NVC designed the research. YY and YD contributed analytic tools and performed empirical evaluation. YY, YD and NVC analyzed the results. YY ,YD, and NVC wrote the paper. The authors declare no conflict of interest. NVC is the corresponding author: nchawla@nd.edu.

\section*{Additional Information}
Competing financial interests: The authors declare no competing financial interests.

\begin{table}[htp]
\caption{Predict Future Degree Centrality. We solve the future degree centrality prediction problem (\textcolor{blue}{Supporting Information, 3}) using supervised learning method. The five NPP positions (Figure~\ref{fig_tpp_example}) census contributes to our {\bf NPP} method for prediction. {\bf PA} (preferential attachment) method just includes the degree centrality feature, and {\bf TC} (triadic closure) method includes the position $3$ as feature. The method labeled {\bf All} includes existing centrality measures (\textcolor{blue}{ Supporting Information, 3.1}). The supervised learning task is to predict whether a new arriving node will become a important node or a non important node (determined by its degree centrality, see \textcolor{blue}{Supporting Information, 3.2}) in future. The experiment settings are provided in \textcolor{blue}{Supporting Information, 3.2}.} % title of Table
\label{performance_metric1}
\centering % used for centering table
\resizebox{3.3in}{!}{
\begin{tabular}{|c|p{1cm}|p{1cm}|p{1cm}|p{1cm}|p{1cm}|p{1cm}|p{1cm}|p{1cm}|} 
\hline
{} & \multicolumn{4}{c|}{AUROC} & \multicolumn{4}{c|}{AUPR}\\
\cline{2-9}
{Datasets} & {PA} & {TC} & All & {NPP} & {PA} & {TC} & All & {NPP}\\
\hline
Condmat & 0.85 & {0.72} & 0.85 & {\bf 0.86} & 0.68 & 0.42 & 0.71 & {\bf 0.72} \\ \hline
DBLP & 0.79 & {0.83} & 0.72 &{\bf 0.85} & 0.27 & 0.34 & 0.19 & {\bf 0.36} \\ \hline
Enron & 0.71 & {0.55} & 0.70 & {\bf 0.72} & 0.43 & 0.18 & 0.51 & {\bf 0.52} \\ \hline
Facebook & {\bf 0.81} & {0.78} & 0.74& {\bf 0.81} & 0.42 & 0.32 & 0.42 & {\bf 0.45} \\ \hline
\end{tabular}
}
\end{table}

% that's all folks
\end{document}